\documentclass[11pt]{article}
\usepackage{pst-all}
\usepackage{pstricks}
\usepackage{pstcol,pst-fill,pst-grad}
\usepackage{amsfonts}
\usepackage{graphicx}
\usepackage{amsmath}

\makeatletter \@addtoreset{equation}{section}

\newcommand{\be}{\begin{equation}}
\newcommand{\ee}{\end{equation}}
\newcommand{\bea}{\begin{eqnarray}}
\newcommand{\eea}{\end{eqnarray}}
\begin{document}
\date{}
\title{
\textbf{  Graph Theory and  Qubit  Information Systems  of   Extremal   Black Branes   }\\
\textbf{   } }
\author{ Adil Belhaj$^{1}$, Moulay Brahim Sedra$^{2,3}$, Antonio Segui$^{4}$
\hspace*{-8pt} \\
\\
{\small $^{1}$D\'epartement de Physique, Facult\'e
Polydisciplinaire, Universit\'e Sultan Moulay Slimane}\\{ \small
B\'eni Mellal, Morocco }
\\ {\small $^{2}$  LHESIR,   D\'{e}partement de Physique, Facult\'{e}
des Sciences, Universit\'{e} Ibn Tofail }\\{ \small K\'{e}nitra,
Morocco}\\
 {\small $^{3}$ Universit\'e Mohammed Premier,
  Ecole Nationale des Sciences Appliqu\'ees}\\{ \small BP : 3, Ajdir, 32003, Al Hoceima, Morocco} \\{\small $^{4}$
 Departamento de F\'isica Te\'orica, Universidad de Zaragoza, E-50009-Zaragoza,
 Spain.} }  \maketitle

\begin{abstract}

Using  graph theory based on  Adinkras, we consider once again the
study of extremal black  branes  in the framework of quantum
information. More precisely, we  propose   a  one to one
correspondence between qubit systems,  Adinkras  and certain
extremal  black branes obtained from
 type IIA superstring  compactified on $T^n$.  We accordingly  interpret the real
Hodge diagram of $T^n$   as  the  geometry  of  a   class of
Adinkras formed by  $2^n$ bosonic
 nodes representing   $n$ qubits. In this graphic  representation, each node encodes  information
on the qubit quantum  states and  the charges of  the extremal
 black  branes  built on  $T^n$. The correspondence is  generalized to  $n$ superqubits associated with
odd and even geometries on   the  real  supermanifold $T^{n|n}$.
Using a combinatorial computation,  general expressions  describing
the number of the  bosonic  and  the fermionic states are obtained.
\end{abstract}


\thispagestyle{empty}

\newpage \setcounter{page}{1} \newpage

\section{Introduction}

 Recently, extremal black   branes in arbitrary dimensions  have been
 investigated using different methods in the framework of  string
 theory and  related topics including
M-theory  compactified on  manifolds   having  special holonomy
groups \cite{1,2,3,4}. These black objects have been studied using
the so called attractor mechanism \cite{40,5,6}.  In this way, the
scalar fields can be fixed in terms of  the  black brane charges.
This analysis can  be done  in terms of an effective potential
depending on the charges and the stringy    moduli obtained from the
compactification of higher dimensional theories. Extremising the
potential with respect to such moduli, the minimum generates the
scalar  fixed values. Moreover, the corresponding entropy functions
have been computed using   the U-duality symmetry acting on the
invariant black brane  charges  of the compactified theories.  In
this regard, the
 Calabi-Yau compactifications have been explored to  produce  several
 interesting results  dealing with  the   black objects in higher
  dimensional  theories including  string theory
  \cite{ 60,61,62,5,6,7,8,9,10,11,12,13,14}.\\
\par
 More recently,  a connection  with   quantum information
has  been proposed using  the qubit formalism\cite{ 140,141,
15,16,17,181}. It is recalled that the qubit is  the building piece
of quantum information theory. In fact, a possible interplay between
 the STU black holes  having  eight charges  and three qubits have
been given in \cite{190,191}. The analysis has been extended to
include superqubits using supersymmetric approach based on the
theory of Lie superalgebras. In particular,  the $osp(2|1)$ Lie
superalgebra has been explored to deal with  the physics of the
superqubits\cite{192,193,18}.
\\
\par
In this paper,
 we reconsider the study of the
extremal black  branes  in the framework of quantum information
using  graph theory based on the so called Adinkras \cite{22,23}.
More precisely, we establish  a  one to one correspondence between
qubit systems, Adinkras  and extremal  black branes   embedded in
maximally supersymmetric  supergravity obtained from a low  energy
limit  of
 type IIA superstring  compactified  on  $T^n$.  We accordingly  interpret the real
Hodge diagram of $T^n$   as  the  geometry  of  a   class of
Adinkras involving   $2^n$
 bosonic nodes representing   $n$ qubits. In this graphic  representation, each node encodes  information
on  the qubit quantum  states and  the charges of  the extremal
 black  branes constructed from the compactification  of type IIA superstring   on  $T^n$. The correspondence is
 generalized to  $n$ superqubits associated with
odd and even geometries on   the  real  supermanifold $T^{n|n}$.
Using combinatorial computation,  general expressions  describing
the number of bosonic  and fermionic states are obtained. Then,
illustrated models  are given.  It is worth noting a that another  nice connection between cohomology of extra
dimensions and qubit systems  have been also given in \cite{181,182} \\
\par

The organization of the paper  is  as follows. In section 2,  we
reconsider the study of the  extremal black   branes in  type IIA
superstring  compactified  on  $T^n$. Section 3 concerns  an
elaboration of a correspondence between qubit systems, Andinkras and
the  real Hodge diagrams  of $T^n$. In section 4, we give a graphic
representation of the  extremal black branes using qubit systems
based on Adinkras. The generalization  to   superqubit  systems is
given in section 5. The analysis is  done   in terms of   the
  real supermanifold $T^{n|n}$. The last section is devoted to  the
conclusion and open questions.

\section{ Extremal black   branes   in  string theory}
In this section,  we reconsider the discussion of  the extremal
black  branes  in string theory  compactified  on  $n$-dimensional
compact manifolds $X^n$. It has been shown that  the  black
$p$-objects can  be produced by a system of $(p+k)$-branes wrapping
the $k$-cycles of $X^n$. In such a  compactification,   it has has
been shown that the near horizon geometries of  the extremal black
$p$-brane are defined by the product of Ads spaces and spheres
\begin{equation}
 Ads_{p+2}\times S^{8-n-p}.
\end{equation}
It is obvious  to see that  the  integers  $n$ and $p$  should
satisfy
\begin{equation}
1 \leq n, \quad  2 \leq 8-n-p.
\end{equation}
The generalization of such geometries  to investigate  intersecting
attractors has been also given in \cite{14}.

In higher dimensional theories, one may classify the the black
$p$-brane solutions  using  the extended electric/magnetic duality
connecting   a $p$-dimensional electrical black   brane to a
$q$-dimensional magnetic one  via the following relation
\begin{equation}
\label{duality} p+q=6-n.
\end{equation}
The solution of this equation provides three different black objects
organized  in terms of     the values  of the charge couple $(p,q)$.
Indeed,  they are given by
\begin{itemize}
\item $(p,q)=(0,6-n)$,  associated with   the electrical charged   black
holes,
\item $(p,q)=(3-\frac{n}{2},3-\frac{n}{2})$ ($n$ even), corresponding to    the dyonic black
branes,  having   both the  electric and magnetic charges,
\item $(p,q)\neq(0,6-n)$, where  $(p,q)\neq (3-\frac{n}{2},3-\frac{n}{2})$,  describing   the  magnetic  black
branes.
\end{itemize}
It is  noted that   $p=0$  describes    the  electric charged black
holes in $10-n$ dimensions.
 Their  dual magnetic are  black  $6-n$-branes.
These  black  objects  carry charges corresponding to   the gauge
invariant field strengths ($F=dA$) of  the corresponding maximally
supersymmetric  supergravity theory. It is worth recalling that even
dimensional space-times produce dyonic
 black  branes. The charges of these objects can fix the value of
 the  dilaton using the attractor mechanism as reported in  \cite{13}.

Before examining    a particular supergravity theory solution, we
should recall that the   black object charges   depend on the choice
of the internal space. It turns out that,  each compactification
involves a Hodge diagram  carrying  not only geometric information
but also physical data  on the  black brane solutions.

In the following sections, we will be  concerned with  the
compactification of type IIA superstring theory on the
$n$-dimensional tori $T^n$.  This compactification   may produce the
black $p$-brane configurations in $10-n$ dimensional maximally
supersymmetric  supergravity coupled to abelian gauge symmetries
associated with the  { \bf NS-NS} and {\bf R-R} bosonic fields of
various ranks. In this way, the black objects  can be constructed
using  the following brane configurations
\begin{equation*}
\mbox{D0-branes}, \quad   \mbox{F-strings}, \quad \mbox{D2-branes},
\quad \mbox{D4-branes},  \quad \mbox{NS-5branes}, \quad
\mbox{D6-branes}.
\end{equation*}
As in the case of    the Calabi-Yau manifold,  the toroidal
compactification involves a  real Hodge diagram  playing   a
primordial   role in the elaboration  of the  type IIA superstring
charge   spectrum in $10-n$ dimensions.  Roughly speaking, $T^n$ is
a flat  compact space  which can be defined using different ways.
One of them is to use the trivial fibration of $n$ circles modeled
by the following  orbital  relations
\begin{equation}
x_i\equiv x_i+1,  \qquad i=1,\ldots,n.
\end{equation}
To bluid the real Hodge diagrams, it  is convenient to introduce a
binary number notation $h^{e_1,\ldots,e_n}$ playing a similar role
as the Hodge numbers appearing in  the cohomology  of complex
Calabi-Yau geometries. More precisely,  this number is associated
with the following differential form
\begin{equation}
 h^{e_1,\ldots,e_n}\longrightarrow \bigwedge_{\ell=1}^{n}(\overline{{e_\ell}}+e_\ell
dx_\ell),
\end{equation}
 where  $e_\ell$ is a binary number taking either $0$ or $1$, and where  $\overline{{e_\ell}}$ is  its
 conjugate. It is obvious to see   that  $\prod_{\ell=1}^{n}(\overline{{e_\ell}}+e_\ell
 dx_\ell)$ is  a real differential form  of degree $k$ which  can be written as follows
\begin{equation}
  k=\sum_{\ell=1}^ne_{\ell}.
\end{equation}
By using  the  Poincar\'e  duality,  this $k$-form is dual to a
$k$-cycle embedded in $T^n$ on which type IIA branes could  wrap to
produce black objects in $10-n$ dimensions.  By inverting the order
of the   Hodge diagram corresponding to  the Calabi-Yau manifolds,
we can give a graphic representation  of the  cohomology space  of
$T^n $ in terms of the   real real Hodge diagrams. For simplicity
reason, we illustrate   the  $n=2$ and $n=3$ cases
\[
\begin{tabular}{|l|l|l|}
\hline
$n=2$ & $%
\begin{tabular}{lll}
& $h^{1,1}$ &  \\
$h^{1,0}$ &  & $h^{0,1}$ \\
& $h^{0,0}$ &
\end{tabular}%
$ & $%
\begin{tabular}{lll}
& $1$ &  \\
$1$ &  & $1$ \\
& $1$ &
\end{tabular}%
$ \\ \hline
$n=3$ & $%
\begin{tabular}{lllll}
&  & $h^{1,1,1}$ &  &  \\
& $h^{1,1,0}$ & $h^{1,0,1}$ & $h^{0,1,1}$ &  \\
&$h^{1,0,0}$ & $h^{0,1,0}$   & $h^{0,0,1}$ \\
 &  & $h^{0,0,0}$ &  &
\end{tabular}%
$ & $%
\begin{tabular}{lllll}
&  & $1$ &  &  \\
& $1$ & $1$ & $1$ &  \\
&$1$ & $1$   & $1$ \\
 &  & $1$ &  &
\end{tabular}
$ \\ \hline
\end{tabular}%
\]
\def\m#1{\makebox[10pt]{$#1$}}
 It is noted that the general  real Hodge  diagram associated with $T^n$ can be
constructed by using the  above mentioned notation and terminology.
It encodes all possible non trivial  cycles of $T^n$   describing
its geometric data including the  size and the shape parameters.

A close inspection shows that there is  a  striking resemblance
between the real Hodge diagram of $T^n$ and    graph theory of  a
particular class of Adinkras. This  leads to a nice correspondence
between the nodes of such a class of Adinkras  and  the cycles
involved in the determination of   the black brane charges in type
IIA superstring on $T^n$. On the other hand,  these Adinkras  have
been explored to give a graphic representation of qubit systems.
Naturally, there should be a relation  between all  these issues.
This question will be addressed in the following sections.

\section{Qubit systems, Andinkras  and real Hodge diagrams}
In this section, we would like  to elaborate  a  link between
qubits, Adinkras and  the real Hodge diagrams of $T^n$. This
correspondence will be explored to engineer graphically the extremal
black brane geometries using qubit systems. It is recalled  that the
qubit is a primordial piece in quantum
 information  theory  which  have been extensively investigated using
  different physical and  mathematical approaches \cite{19,20,21}.
  It is a two configuration system which can be
 associated, for instance, with the electron  in the hydrogen atom.
 The general state of  a single  qubit is usually given by the
 Dirac notation as follows
\begin{equation}
|\psi\rangle=c_0|0\rangle+c_1 |1\rangle
\end{equation}
 where $c_i$  are complex  coefficients  satisfying the normalization
condition
\begin{equation}
|c_0|^2+|c_1 |^2=1.
\end{equation}
  This equation can be  interpreted  geometrically in terms of the  so called Bloch
sphere. The two quits are four configuration systems. In this case,
the general state takes the following form
\begin{equation}
|\psi\rangle=c_{00}|00\rangle+c_{10}
|10\rangle+c_{01}|01\rangle+c_{11} |11\rangle
\end{equation}
where $c_{ij}$  are complex numbers satisfying the normalization
condition
\begin{equation}
|c_{00}|^2+|c_{10}|^2+|c_{01}|^2+|c_{11}|^2=1,
\end{equation}
describing a   three dimensional complex projective space $CP^3$
generalizing the Bloch sphere. This  analysis can be extended  to
$n$ qubits associated with  $2^n$ configuration states using the
above binary notation.

  It is observed  that
the  qubit systems can be represented by   diagrams having a strong
resemblance with
 a particular class of  Adinkras.  For later, we refer to them as bosonic graphs.
The Adinkras have been introduced in the study of supersymmetric
representation theory \cite{22,23,24,25,26,27,28}. As   the usual
diagram representation, Adinkras are formed by nodes and   lines. It
has been shown that  there are  various  classes  which have been
explored  in the classification of supersymmetric  theories. These
diagrams contain  bosonic and fermionic nodes like Dynkin diagrams
of Lie superalgebras. A particular graph  is called regular  formed
by $2^n$ nodes connected  with
 $n$ colored lines.  In fact,  this   graph  has been explored to represent $n$-qubits.   More
specifically, to each node,  one  associates  a   state  of the
qubit. To be precise, the  correspondence can be formulated as
follows
\begin{eqnarray}
\mbox{node} &\longrightarrow &  \mbox{state} \\
\mbox{ number of colors} & \longrightarrow & \mbox{ number of
qubits}
\end{eqnarray}
To illustrate,  we  present  two bosonic graphs associated with
$n=2$ and $n=3$. The  first example concerns  $n=2$ qubits and it is
given by

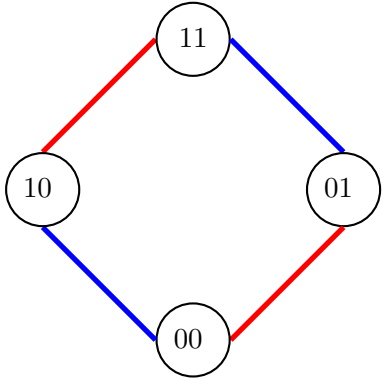
\vspace{0.5cm}

\begin{center}
\begin{pspicture}(-5,5)(3,3)
\begin{pspicture}(3,-3)(5,3)
\psline[linewidth=2pt,linecolor=red](-2,0.5)(-0.5,2)
\pscircle(0,2){0.5}
\psline[linewidth=2pt,linecolor=blue](0.5,2)(2,0.5)
\pscircle(2,0){0.5}
\psline[linewidth=2pt,linecolor=red](2,-0.5)(0.5,-2)
\pscircle(0,-2){0.5}
\psline[linewidth=2pt,linecolor=blue](-0.5,-2)(-2,-0.5)
\pscircle(-2,0){0.5}

\uput{-0.1}[u](0,2 ){11} \uput{-0.1}[u](2,0 ){01 }
\uput{-0.1}[u](0,-2){00 } \uput{-0.1}[u](-2,0 ){10 }

\end{pspicture}
\end{pspicture}
\end{center} \vspace{2.5cm}
\newpage
The second example describes $n=3$ qubits and it is represented  by
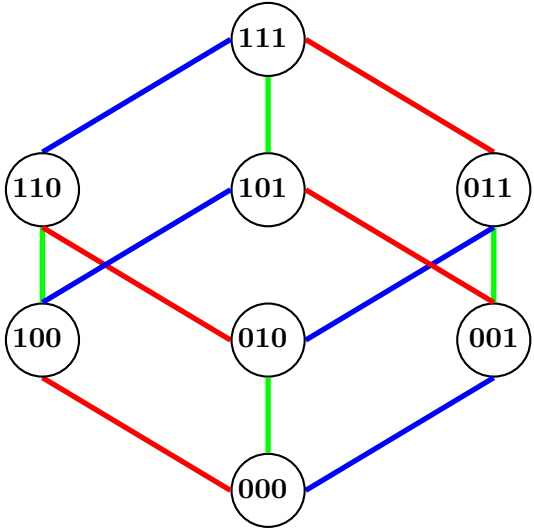
\vspace{2cm}
\begin{center}
\begin{pspicture}(-5,5)(3,3)
\begin{pspicture}(3,-3)(5,3)
\pscircle(0,3){0.5}
\psline[linewidth=2pt,linecolor=red](0.5,3)(3,1.5)
\pscircle(3,1){0.5}
\psline[linewidth=2pt,linecolor=green](3,0.5)(3,-0.5)
\pscircle(3,-1){0.5}
\psline[linewidth=2pt,linecolor=blue](3,-1.5)(0.5,-3)
\pscircle(0,-3){0.5}
\psline[linewidth=2pt,linecolor=red](-0.5,-3)(-3,-1.5)
\pscircle(-3,-1){0.5}
\psline[linewidth=2pt,linecolor=green](-3,-0.5)(-3,0.5)
\pscircle(0,1){0.5}
\psline[linewidth=2pt,linecolor=blue](-3,1.5)(-0.5,3)
\pscircle(-3,1){0.5}
\psline[linewidth=2pt,linecolor=red](-3,0.5)(-0.5,-1)
\pscircle(0,-1){0.5}
\psline[linewidth=2pt,linecolor=blue](0.5,-1)(3,0.5)
\psline[linewidth=2pt,linecolor=blue](-0.5,1)(-3,-0.5)
\psline[linewidth=2pt,linecolor=red](0.5,1)(3,-0.5)
\psline[linewidth=2pt,linecolor=green](0,1.5)(0,2.5)
\psline[linewidth=2pt,linecolor=green](0,-1.5)(0,-2.5)

\uput{-0.1}[u](0,3 ){{\bf 111 }} \uput{-0.1}[u](3,1 ){{\bf 011 }}
\uput{-0.1}[u](3,-1 ){{\bf 001}} \uput{-0.1}[u](0,-3 ){{\bf 000 }}
\uput{-0.1}[u](-3,-1 ){{\bf 100 }} \uput{-0.1}[u](-3,1 ){{\bf 110 }}
\uput{-0.1}[u](0,1 ){{\bf 101 }} \uput{-0.1}[u](0,-1 ){{\bf 010 }}

\end{pspicture}
\end{pspicture}

\end{center}

\vspace{4cm} An inspection shows   that  the real Hodge diagrams of
$T^n$ can be interpreted as a regular  Andikra.  To each node, we
associate then a single cycle in $T^n$.  As we will see shortly in
an example, this link   is  as follows
\begin{equation}
h^{e_1,\ldots,e_n}  \longrightarrow \mbox{node}=(e_1,\ldots,e_n).
\end{equation}
The number of the nodes  in the  $k$-level is  exactly the
combinatorial number
\begin{equation}
\mbox{nbr}(k\mbox{-level nodes)}= C^k_n
\end{equation}
indicating  also    the number of  the $k$-cycle in $T^n$.  The
total number
  of the cycles  is  identified with   the total   number of the nodes. This number is
  given by
\begin{equation}
\mbox{nbr(cycles)}=\sum_{k=0}^{n} C^k_n=2^n.
\end{equation}
On the basis of the above link, the  cycles  in $T^n$  should be
associated with the  states defining   the  $n$ qubit systems. In
this way, a   quantum state is interpreted as the Poincar\'e dual of
the
 real  homology cycle on which branes can wrap to produce a
black object in $10-n$ dimensional type IIA superstring. We expect
that
 the Adinkras should encode information  on  the black  brane physics in  higher dimensional supersymmetric
 supergravity  theories.
   This
may offer a new take on the graphical representation of such a
physics  using techniques  of  graph theory.  The following parts
concern  graphic representations  of  the extremal black branes
using qubit and Adinkra notions.

\section{ Extremal black branes  and Andinkra representation of qubit systems}

In  this section,  we   examine  the correspondence between the
extremal  black $p$-branes  in    maximally supersymmetric
supergravity and qubit systems using  a graph theory based on
bosonic  Adinkras. To understand how such a surprising connection
could be true, we first consider  the case of  the elliptic curve.
Then,  we give a general picture which may  appear in lower
dimensional  theories.

\subsection{  The elliptic curve compactification}
 As mentioned, we  start   by  discussing  the  eight dimensional extremal black branes.
  In particular, we
will give  two explicit models corresponding to  $p=0$ and $p=4$.
These models  can be obtained from the compactification of type IIA
superstring on $T^{2}$, considered as a trivial fibration of two
circles:  $ S^1\times S^1$.   The compactification of  the massless
bosonic ten dimensional  type IIA superstring fields
\begin{equation}
{NS-NS}:G_{MN},\;\;B_{MN},\;\;\phi \qquad {R-R} :A_{M},\;\;C_{MNK}
\quad  M,N,K=0,\dots,9
\end{equation}
gives   an  eight dimensional spectrum  encoding  also information
on  the black brane charges.  This spectrum,  which can be
alternatively obtained from the compactification of M-theory on
$T^{3}$ with  the $ \textsc{ SL(3,R)}\times \textsc{ SL(2,R)}$
U-duality group \cite{29},    contains   a graviton, three 2-form
gauge fields, six vector gauge  fields, one self dual 3-form gauge
field and seven scalar fields. The  associated scalar manifold reads
as
\begin{equation}
\frac{\textsc {SL(3,R)}}{\textsc {SO(3)}}\times \frac{\textsc{
SL(2,R)}}{\textsc {SO(2)}}.
\end{equation}

Motivated by the study of the attractors on the K3 surface in type
IIA superstring, another factorization of such a moduli  space has
been  given in  \cite{12}. In type IIA six dimensions obtained from
the compactification of the K3 surface, the moduli space contains
two factors associated with two possible brane charges given by
black strings and holes. According to \cite{12}, the separation of
the extremal black brane charges in eight dimensions  provides   a
possible factorization of the above  scalar manifold given by
\begin{equation}
\label{factor}
\frac{ \textsc{SO(2,2)}}{ \textsc{SO(2)}\times  \textsc{SO(2)}}\times \frac{%
 \textsc{SO(2,1)}}{ \textsc{SO(2)}}\times  \textsc{SO(1,1)}.
\end{equation}
associated with three possible black brane charges. In this  eight
dimensional $N = 2$ supergravity,  the extremal near-horizon
geometries of the black  $p$-branes  take  the following form
\cite{14}
\begin{equation}
Ads_{p+2}\times S^{6-p}.
\end{equation}
Indeed, they are classified into three categories:
\begin{enumerate}
  \item  $p = 0$  is  associated  with
$AdS_2 \times  S^6$  which describes  the near-horizon geometry of
eight dimensional  electric charged black holes. Their dual magnetic
objects  are black 4-branes having   the $AdS_6 \times  S^2$
near-horizon geometry.  They   are  charged under six gauge field
strengths $F_i = dA_i (i = 1,\ldots, 6)$ corresponding to the
following gauge symmetry
\begin{equation}
G=U(1)^4\times  U(1)^2.
\end{equation}
  \item  $p = 1$ corresponds to  the  extremal black strings with  the  $AdS_3 \times  S^5$   near-horizon geometry.
   The magnetic dual horizon
geometry  is  $AdS_5 \times  S^3$   associated with  the  black
3-branes.
  \item  $p = 2 $  describes   a dyonic black 2-brane having  $AdS_4\times  S^4$
near-horizon geometry. This black object  can   carry  both electric
and magnetic charges.
\end{enumerate}

To see how the  interplay    works,   we  discuss  the case of the
black hole solution  where   the Kaluza Kein states associated with
the  symmetry  $U(1)^2_{bh}$ are turned off. More precisely, we
construct a black hole from a system of $\{\mbox{D0-branes}, \mbox{
F-strings},\mbox{D2-branes}\}$ wrapping  appropriate cycles of the
elliptic curve $T^2$.  In this way,  the  dictionary between these
cycles,  four node Andinkra  and two qubits
  may offer a new   way to investigate  eight dimensional  black holes. Indeed, the $\frac{SO(2,2)}{SO(2) \times SO(2)
  }$ appearing in (\ref{factor})
should correspond to
 a black hole with    4 electric   charges
 which  is exactly  the entries appearing in  the real   Hodge diagram of the elliptic curve $T^2$.
Its   brane   representation can be  organized  in the following
brane
 diagram
\begin{equation}
  {\arraycolsep=2pt
  \begin{array}{*{5}{c}}
    &&\m1&& \\ &&&& \\ \m1&&&&\m{1} \\
    &&&& \\ &&\m1&&
  \end{array}} \;=\;{\arraycolsep=2pt
  \begin{array}{*{5}{c}}
    && D2&& \\ &&&& \\ \mbox{F-string}&&&&\mbox{F-string}\\
    &&&& \\ &&D0&&
  \end{array}}
\end{equation}
shearing similarities  with the real Hodge diagram  of $T^2$. Using
equation (\ref{duality}) for $n=2$, we can give a graphic
representation for the magnetic black 4-brane. It is given by the
following brane diagram
\begin{equation}
  {\arraycolsep=2pt
  \begin{array}{*{5}{c}}
    &&\m1&& \\ &&&& \\ \m1&&&&\m{1} \\
    &&&& \\ &&\m1&&
  \end{array}} \;=\;{\arraycolsep=2pt
  \begin{array}{*{5}{c}}
    && D6&& \\ &&&& \\ \mbox{NS5-brane}&&&&\mbox{NS5-brane}\\
    &&&& \\ &&D4&&
  \end{array}}
\end{equation}

Using the previous notations, we propose the  correspondence given
in table 1.

\begin{table}[!ht]
\begin{center}
\begin{tabular}{|c|c|c|c|c|}
\hline $T^2$ & Adinkra &  qubit system &  Black hole system & Black
4-brane system
 \\ \hline 1  & (00)& $ |00\rangle$&  D0-brane& D4-brane \\ \hline   $dx_1$ & (01) & $|01\rangle$ & F-string
 &NS5-brane
 \\ \hline
 $dx_2$ & (10) & $|10\rangle$ & F-string & NS5-brane
\\
 \hline
$dx_1dx_2$  & (11) & $|11\rangle$ & D2-brane & D6-brane
\\ \hline
\end{tabular}
\end{center}
\label{tab2} \caption{This table gives  the correspondence between
the  eight dimensional black  branes, Adinkra and qubit systems.}
\end{table}

The  same analysis  can be done for $p=1$ and  $p=2$  but with some
differences required by the  T-duality between D2-branes and
D3-branes which  has been extensively studied in connection with
string theory compactification \cite{290,291}. The corresponding
brane diagrams should involve D3-branes.

\subsection{ Lower dimensional cases}

Here, we would like to extend the above results to higher orders of
Adinkras.  Particulary,  we can do something similar for the more
general case associated with ($n>2$).  The corresponding
compactification has been extensively studied  to  generate  lower
dimensional type IIA superstring.  In this way, the  Adinkras
 will be based on
 the polyvalent geometry  in which the  nodes  are connected with   more than two
other ones
 as shown in the  previous section  associated with  the trivalent geometry. The latter
 generalizes
the bivalent one appearing in the case of the elliptic curve $T^2$.
Based on this observation,  we believe that  the physics  of the
extremal black
 branes    in  type IIA superstring  on  $T^n$ shears similarities with  $n$
 qubits. Before giving a general picture, we illustrate the  case of  $T^3$.
 The  corresponding  compactification produces  seven dimensional  extremal black
 branes.  Their extremal
near-horizon geometries  are given by
\begin{equation}
Ads_{p+2}\times S^{5-p}.
\end{equation}
These asymptotically flat, static and spherical solutions are
classified by the following fundamental solutions
\begin{enumerate}
\item  seven dimensional  black holes dual to black 3-branes,
\item seven dimensional   black strings dual to black 2-branes.
\end{enumerate}
The seven dimensional  correspondence, in  the presence of the
D3-brane, can be illustrated in table 2.
\begin{table}[!ht]
\begin{center}
\begin{tabular}{|c|c|c|c|}
\hline $T^3$ & Adinkra &  qubit system &  Black hole system
 \\ \hline 1  & (000)& $ |000\rangle$&  D0-brane\\ \hline   $dx_1$ & (001) & $|001\rangle$ & F-string
 \\ \hline
  $dx_2$ & (010) & $|010\rangle$ & F-string \\ \hline

 $dx_3$ & (100) & $|100\rangle$ & F-string
\\
 \hline
$dx_1dx_2$  & (011) & $|011\rangle$ & D2-brane
\\ \hline
$dx_2dx_3$  & (110) & $|110\rangle$ & D2-brane
\\ \hline

$dx_1dx_3$  & (101) & $|101\rangle$ & D2-brane
\\ \hline
$dx_1dx_2dx_3$  & (111) & $|111\rangle$ & D3-brane (T-dual of
D2-brane)
\\ \hline
\end{tabular}
\end{center}
\label{tab2} \caption{This table gives  the correspondence between
eight dimensional black hole, Adinkra and qubit systems.}
\end{table}

The same analysis  can be done for  the extremal black hole in type
IIA  superstring on $T^n$.   In this case,   the  near horizon
geometry  of the   extremal black holes  reduces to
\begin{equation} Ads_{2}\times S^{8-n}.
\end{equation}
Using the fact that  the real Hodge diagram of $T^n$ can  encode the
information on the black brane charges and $n$ qubit systems,   we
can give  a  possible brane configuration producing extremal  black
branes in $10-n$ dimensions,  obtained from  the compactification of
type IIA superstring on $T^n$.  Similar discussion   be also done
for the extremal black  branes.  For illustration,  the general
picture for the black holes  can be summarized in table 3.
\begin{table}[!ht]
\begin{center}
\begin{tabular}{|c|c|c|c|}
\hline $T^n$ & Adinkra &  qubit system &  Black hole system
 \\ \hline $\prod\limits_{\ell=1}^{n}(\overline{{e_\ell}}+e_\ell dx_\ell)$
  & $(e_1,\ldots,e_n)$& $ |e_1,\ldots,e_n\rangle$&
  $k$-brane\\ \hline
\end{tabular}
\end{center}
\label{tab2} \caption{This table gives  the correspondence between
 $10$-$n$ dimensinal black holes, regular Adinkras and  and $n$ qubit systems.}
\end{table}


\section{ Odd and even geometries on $T^{n|n}$ and  superqubits}
Having discussed the bosonic qubits, it is  very natural to consider
superqubit systems and supersymmetric Adinkras in the above
established correspondence.  In fact,   superqubits have been
  investigated in connection with Lie supersymmetries\cite{16,17,18}.  Roughly speaking,
a superqubit can take  three values: 0 or 1 or $\bullet$. In fact, 0
and 1 are bosonic while  $\bullet$  is fermionic.  In this section,
we investigate the superqubits in  the framework of the toroidal
compactification with  bosonic and fermionic coordinates. In
particular, we explore the superform cohomology of a particular real
supermanifold $T^{n|m}$ equipped with $n$ bosonic coordinates $x_i$
and $m$ fermionic coordinates $\theta_\alpha$. It is recalled that
 the supermanifolds have been extensively  studied in
 string theory and related topics including
mirror  geometries \cite{30,31,32}.\\
In what follows, we consider a special geometry  where $n=m$
corresponding to   the   real supermanifold  $T^{n|n}$ and  make
contact with superqubits.  More precisely,  we would like  to
elaborate  the  real Hodge diagram of $T^{n|n}$.  Up some
assumptions specified later on, we show that a part  of   this
extended  real  Hodge diagram   can be explored to represent $n$
superqubits using the supersymmetric Adinkras involving bosonic and
fermionic nodes. In fact, these supersymmetric real Hodge diagrams
can be built using the extension of  the   above binary notation.
For simplicity reason  and keeping the contact with the previous
section, we use  the  notation
$h^{e_1,\ldots,e_n|\alpha_1,\ldots,\alpha_n}$ associated with  the
following differential superform
\begin{equation}
 h^{e_1,\ldots,e_n|\alpha_1,\ldots,\alpha_n}
 \longrightarrow  \prod_{\ell=1}^{n}(\overline{{e_\ell}}+e_\ell
 dx_\ell) \prod_{\alpha=1}^{n}(\overline{{e_\alpha}}+e_\alpha d\theta_\alpha)
\end{equation}
Its degree is given by
\begin{equation}
  d=\sum_{\ell=1}^ne_{\ell}+ \sum_{\alpha=1}^ne_{\alpha}.
\end{equation}
To give a differential geometry representation of   $n$ superqubits,
we postulate the following   constraints on the superform degrees:
\begin{itemize}
\item the lowest degree is zero associated with a bosonic state
\item the highest  degree is $n$ associated either with  a bosonic or
a fermionic state  according to the parity of $n$.
\end{itemize}
More precisely, we will be interested in  the following superforms
on  the real  supermanifold $T^{n|n}$
\begin{equation}
\prod_{\ell=1}^{n}(\overline{{e_\ell}}+e_\ell dx_\ell)
\prod_{\alpha=1}^{n-\ell}(\overline{{e_\alpha}}+e_\alpha
d\theta_\alpha).
\end{equation}
Based on the above assumption, combinatorial calculation reveals
that the total   number of the associated   cycles   is given  by
\begin{equation}
\sum_{\ell=0}^n C_n^l\sum_{k=0}^{n-\ell} C_{n-\ell}^k.
\end{equation}
It is worth noting that if we put $\ell=0$, we recover  the bosonic
case as discussed in section 2.  Using the expression of  the Taylor
series $(1+x)^n$, we can show that this number is exactly the state
number of  $n$ superqubits
\begin{equation}
\sum_{\ell=0}^n C_n^l\sum_{k=0}^{n-\ell} C_{n-\ell}^k=3^n.
\end{equation}
This includes the bosonic and  the fermionic  states.  To get the
state number of each sector, we use the following   splitting
\begin{equation}
3^n= \frac{3^n-1}{2} +\frac{3^n+1}{2}.
\end{equation}
The calculation shows  the following relations
\begin{eqnarray}
\sum_{\ell=even}^n C_n^l\sum_{k=0}^{n-\ell} C_{n-\ell}^k=\frac{3^n+1}{2}\\
\qquad \sum_{\ell=odd}^n C_n^l\sum_{k=0}^{n-\ell}
C_{n-\ell}^k=\frac{3^n-1}{2}.
\end{eqnarray}
The number of  the  bosonic and  the fermionic states can be
identified by exploring the  following formal supersymmetry
structure
\begin{eqnarray*}
BB=B\\ BF=F\\
 FB=F\\
 FF=B
\end{eqnarray*}
where $B$  and  $F$ denote  the  bosonic  and  the fermionic
generators respectively.  An  inspection shows that the number of
the  bosonic states is given
 \begin{eqnarray}
\mbox{Number of bosonic states}=\frac{3^n+1}{2}.
\end{eqnarray}
Similarly, the number of  fermionic states  reads as
\begin{eqnarray}
\mbox{Number of fermionic states}=\frac{3^n-1}{2}.
\end{eqnarray}
In  connection with graph theory,  the corresponding Adinkras should
be formed by $\frac{3^n+1}{2}$ bosonic nodes and  $\frac{3^n-1}{2}$
fermonic nodes associated with even and odd geometries on $T^{n|n}$
respectively.  In this way, the link  can be put as follows
\begin{eqnarray}
\mbox{Bosonic nodes} &\longrightarrow &  \mbox{Bosonic forms} \\
\mbox{ Fermonioc nodes} & \longrightarrow & \mbox{ Fermionic forms}.
\end{eqnarray}
To illustrate this analysis, we examine the  real  supermanifold
$T^{2|2}$ associated with two superqubits. In particular,  it is
obvious to see that   the bosonic  states correspond to the
following differential form
\begin{equation}
1,\quad dx_1,\quad dx_2,\quad dx_1dx_2, \quad d\theta_1 d\theta_2.
\end{equation}
The number of the corresponding  bosonic states is
$\frac{3^2+1}{2}=5$. Similarity, we can get  the fermionic  states
associated with
\begin{equation}
d\theta_1,\quad d\theta_2,\quad dx_1d\theta_2, \quad d\theta_1 dx_2.
\end{equation}
The   number  is  $\frac{3^2-1}{2}=4$. Motivated by the existence
the supermanifolds and brane geometries, the corresponding Adinkras
and brane charges will be  represented in table 4. The table will
include superbranes  living in supermanifolds.

\begin{table}[!ht]
\begin{center}
\begin{tabular}{|c|c|c|c|}
\hline $T^{2|2}$ & Adinkra &  qubit system &  black hole system
 \\ \hline 1  & $(00|00)$& $ |00|00\rangle$& Bosnic  D0-brane\\ \hline   $dx_1$ & $(10|00)$ & $|10|00\rangle$ &
 Bosonic  F-string
 \\ \hline
  $dx_2$ & $(01|00)$ & $|01|00\rangle$ &  Bosnic F-string \\ \hline

 $dx_1dx_2$ & $(11|00)$ & $|11|00\rangle$ & Bosonic  D2-brane
\\
 \hline
$d\theta_1$  & $(00|10)$ & $|00|\bullet0\rangle$ & Fermionic
F-string
\\ \hline
$d\theta_2$  & $(01|00)$ & $|01|0\bullet\rangle$ & Fermionic
F-string
\\ \hline

$d\theta_1dx_2$  & $(01|10)$ & $|01|\bullet0\rangle$ & Fermionic
F-string
\\ \hline
$d\theta_2dx_1$  & $(10|01)$ & $|10|0\bullet\rangle$ & Fermionic
D2-brane
\\ \hline
$d\theta_1d\theta_2$  & $(00|11)$ & $|00|\bullet\bullet\rangle$ &
Bosnonic D2-brane
\\ \hline
\end{tabular}
\end{center}
\label{tab2} \caption{This table gives  an illustrated model based
on  the real  supermanifold $T^{2|2}$.}
\end{table}
We can construct many additional examples by considering D3-branes
and theirs supersymmetric versions.
\section{ Conclusion and discussions}
In this paper,
 we  have reconsidered the study of the
extremal black  branes  in the framework of quantum information
using  graph theory based on the so called Adinkras.  In particular,
we have elaborated   a  one to one correspondence between qubit
systems, Adinkras and extremal  black  branes   embedded in
maximally supergravity obtained from a low  energy limit  of
 type IIA superstring  compactified  on  $T^n$. It has been  observed that    the  physical states of
$n$ qubit systems can be represented graphically  using  Adinkras.
These graphs  are based on polyvalent geometries appearing in  the
case of  Dynkin diagrams of Lie algebras. In this representation,
the $n$-qubits are associated with the $n$-valent geometry in which
each node is connected with $n$ colored lines. Based on this
observation, the $n$ qubit is represented by a graph of $2^n$
bosonic nodes connected by the colored  $n$ polyvalent geometry.  In
this graphic representation, each node encodes  information on  the
qubit quantum states and the charges of  the extremal
 black  branes  embedded in type IIA superstring on  $T^n$. Then, we
have  proposed a possible generalization to   $n$ superqubits. More
precisely,  we have shown  that these systems can be  associated
with odd and even geometries on the  real supermanifold $T^{n|n}$.
More precisely, the number of the corresponding bosonic and
fermionic states are obtained using  a combinatorial calculation.

Our paper comes up with many open questions  related to quantum
information theory.  In fact, many concepts have been developed in
such a theory including gates, circuits and entanglement states. It
should of be of interest to investigate such concepts using  the
graph theory and  the  physics dealing with black objects. Moreover,
there have been many works trying to connect  black hole with
quantum information \cite{320}. It should be interesting to make
contact with such activities. This will be addressed elsewhere.

 On
Other hand, the analysis presented here might be extended in the
case of the Calabi-Yau manifolds. To speculate on this extension,
let us consider the compactification  of  type IIA superstring on
the K3 surface \cite{321,322}.  It is recalled the the corresponding
complex Hodge diagram reads  as
\[
\begin{tabular}{lllll}
&  & $h^{0,0}$ &  &  \\
& $h^{1,0}$ &  & $h^{0,1}$ &  \\
$h^{2,0}$ &  & $h^{1,1}$ &  & $h^{0,2}$ \\
& $h^{2,1}$ &  & $h^{1,2}$ &  \\
&  & $h^{2,2}$ &  &
\end{tabular}=
\begin{tabular}{lllll}
&  & $1$ &  &  \\
& $0$ &  & $0$ &  \\
$1$ &  & $20$ &  & $1$ \\
& $0$ &  & $0$ &  \\
&  & $1$ &  &
\end{tabular}%
\]
It has been shown that the moduli space of such a compactification
is given by the following factorization
\begin{equation}
\label{modulik3} \frac{SO(4,20)}{SO(4)\times SO(20)}\times SO(1,1).
\end{equation}
It is observed that this  factorization is  linked to   two possible
black object solutions in six dimensions which are given by
\[
\begin{array}{cc}
p=0: &   \text{  corresponding to   the black holes with a near-horizon geometry:  $Ads_{2}\times  S^{4}$}\\
p=1: & \text{ associated with a black string having a  near-horizon
geometry: $Ads_{3} \times S^{3}$}.
\end{array}
\]
In fact, the factor $\frac{SO(4,20)}{SO(4) \times SO(20) }$ is
associated with   24  black hole  charges  identified with  entries
appearing in the  above complex Hodge diagram.  The corresponding
brane realization can be given by
\begin{equation}
  {\arraycolsep=2pt
  \begin{array}{*{5}{c}}
    &&\m1&& \\ &&&& \\ \m1&&\m{20}&&\m{1} \\
    &&&& \\ &&\m1&&
  \end{array}} \;=\;{\arraycolsep=2pt
  \begin{array}{*{5}{c}}
    && D0&& \\ &&&& \\ D2&&D2&&D2\\
    &&&& \\ &&D4&&
  \end{array}}
\end{equation}
At first sight,  the  brane representation could be related to  the
qudit  systems \cite{323}.  However, we believe that  this
connection deserves a deeper study.

 \vspace{1cm}

{\bf Acknowledgements}:   It is a pleasure to thank Prof N. Askour
for discussions.  AS is supported by the Spanish MINECO (grants
FPA2009-09638 and FPA2012-35453) and DGIID-DGA (grant 2011-E24/2).


\begin{thebibliography}{99}
\bibitem{1}
A. Strominger, C. Vafa, {\em  Microscopic Origin of the
Bekenstein-Hawking Entropy}, Phys.Lett. {\bf B379} (1996) 99-104,
{\tt arXiv:hep-th/9601029}.

\bibitem{2} C. Vafa, {\em
 Black Holes and Calabi-Yau Threefolds},
Adv.Theor.Math.Phys.{\bf 2} (1998) 207-218, {\tt hep-th/9711067}.

\bibitem{3}
 J. Maldacena, A.
Strominger, E. Witten, {\em Black Hole Entropy in M-Theory}, JHEP
9712 (1997)002, {\tt arXiv:hep-th/9711053}.


\bibitem{4} R. Ahl Laamara, A. Belhaj, L. B. Drissi, E. H. Saidi, \emph{%
Black Holes in type IIA String on Calabi-Yau Threefolds with Affine
ADE
Geometries and q-Deformed 2D Quiver Gauge Theories}, Nucl. Phys. \textbf{B776%
} (2007) 287-326, \texttt{hep-th/0611289}.
\bibitem{40}
S. Ferrara, R. Kallosh and A. Strominger, N = 2 extremal black
holes, Phys. Rev. D52 (1995) 5412, hep-th/9508072. A. Strominger,
Macroscopic entropy of N = 2 extremal black holes, Phys. Lett. B383,
39 (1996), hep-th/9602111. S. Ferrara and R. Kallosh, Supersymmetry
and attractors, Phys. Rev. D54, 1514 (1996), hep-th/9602136. S.
Ferrara and R. Kallosh, Universality of supersymmetric attractors,
Phys. Rev. D54, 1525 (1996), hep-th/9603090. S. Ferrara, G. W.
Gibbons and R. Kallosh, Black Holes and Critical Points in Moduli
Space, Nucl. Phys. B500 (1997) 75, hep-th/9702103.

\bibitem{5} H. Ooguri, A. Strominger, C. Vafa, \emph{Black Hole Attractors
and the Topological String}, Phys. Rev. \textbf{D70} (2004) 106007, \texttt{%
hep-th/0405146}.

\bibitem{6} S. Ferrara and R. Kallosh, \emph{Supersymmetry and Attractors},
Phys. Rev. \textbf{D54} (1996) 1514, \texttt{hep-th/9602136}.
\bibitem{60}
A. Ceresole, S. Ferrara and A. Marrani, 4d/5d Correspondence for the
Black Hole Potential and its Critical Points, Class. Quant. Grav.
24, 5651 (2007), arXiv:0707.0964 [hep-th].

\bibitem{61}
 S. Bellucci, S. Ferrara, A. Marrani and A. Yeranyan, Mirror Fermat
Calabi-Yau three-folds and Landau-Ginzburg Black Hole Attractors,
Riv. Nuovo Cim. 029, 1 (2006), hep-th/0608091

\bibitem{62}
 K. Saraikin and C. Vafa, Non-supersymmetric black holes and topological strings, Class.
Quant. Grav. 25 (2008) 095007, hep-th/0703214.



\bibitem{7} S. Bellucci, S. Ferrara, R. Kallosh, A. Marrani, \emph{%
Extremal Black Hole and Flux Vacua Attractors}, Lect. Notes Phys.
{\bf 755}(2008)115-191, \texttt{arXiv:0711.4547 [hep-th]}.

\bibitem{8} A. Sen, \emph{Black Hole Entropy Function, Attractors and
Precision Counting of Microstates}, Gen. Rel. Grav. {\bf 40}(2008)2249-2431, \texttt{arXiv:0708.1270 [hep-th]}.

\bibitem{9} A. Dabholkar, \emph{Black hole entropy and attractors}, Class.
Quant. Grav. \textbf{23} (2006) 957-980.

\bibitem{10} P. K. Tripathy, S. P. Trivedi, \emph{Non-supersymmetric
attractors in string theory}, JHEP \textbf{0603} (2006) 022, \texttt{%
hep-th/0511117}.

\bibitem{11} A. Belhaj, L. B. Drissi, E. H. Saidi, A. Segui, \emph{N=2
Supersymmetric Black Attractors in Six and Seven Dimensions}, Nucl. Phys.
\textbf{B796} (2008) 521-580, \texttt{arXiv:0709.0398}.

\bibitem{12}
 R. Ahl
Laamara, M. Asorey, A. Belhaj, A, Segui,  {\em  Extremal Black Brane
Attractors on The Elliptic Curve}, J.Phys. {\bf A43}(2010) 105401,
{\tt arXiv:0907.0093}.

\bibitem{13} A. Belhaj, \emph{On Black Objects in Type IIA Superstring
 Theory on Calabi-Yau
Manifolds}, African Journal  Of Math.  Phys. Vol. {\bf 6}(2008)49-54,
 \texttt{arXiv:0809.1114  [hep-th]}.

\bibitem{14}
  S. Ferrara, A. Marrani, J. F. Morales, H.
Samtleben,  \emph{ Intersecting Attractors}, \texttt{arXiv:0812.0050
[hep-th]}.
\bibitem{140}
 M. J. Duff, String triality, black hole entropy and Cayley's hyperdeterminant, Phys. Rev.
D76 (2007) 025017, hep-th/0601134.
\bibitem{141}
 M. J. Duff and S. Ferrara, Black
hole entropy and quantum information, Lect. Notes Phys. 755 (2008)
93, hep-th/0612036.




\bibitem{15}
L. Borsten, M. J. Duff, P. L\'evay, {\em
 The black-hole/qubit correspondence: an
up-to-date review},  {\tt arXiv:1206.3166}.

\bibitem{16}
L. Borsten, M.J. Duff, A. Marrani, W. Rubens, {\em On the
Black-Hole/Qubit Correspondence},
  Eur.Phys.J.Plus {\bf 126} (2011)
37, {\tt arXiv:1101.3559}.

\bibitem{17} M. J. Duff,
S. Ferrara, {\em Four curious supergravities}, Phys.Rev. {\bf D83}
(2011) 046007,  {\tt arXiv:1010.3173}.

\bibitem{181}
P. L\'evay, {\em  Qubits from extra dimensions},  Phys. Rev. {\bf D
84} (2001) 125020.


\bibitem{190}
L. Borsten, D. Dahanayake, M.J. Duff, W. Rubens, H. Ebrahim,
Freudenthal triple clas sification of three-qubit entanglement,
Phys. Rev. A80 (2009) 032326, arXiv:0812.3322 [quant-ph].
\bibitem{191} L. Borsten, D. Dahanayake, M. J. Duff, A. Marrani, W. Rubens, Four-qubit entanglement
classification from string theory, Phys. Rev. Lett. 105, 100507
(2010), arXiv:1005.4915 [hep-th].



\bibitem{192}
L. Borsten, D. Dahanayake, M.J. Duff, W. Rubens, Superqubits, Phys.
Rev. D81 (2010) 105023, arXiv:0908.0706 [quant-ph].

\bibitem{193} L.
Castellani, P. A. Grassi, L. Sommovigo, Triality Invariance in the
N=2 Superstring, Phys. Lett. B678 (2009) 308, arXiv:0904.2512
[hep-th].


\bibitem{18}

L. Borsten, K. Bradler, M. J. Duff,  {\em Tsirelson's bound and
supersymmetric entangled states},  {\tt arXiv:1206.6934}.



\bibitem{180}
L. Castellani, P. A. Grassi, L. Sommovigo,  {\em   Quantum Computing
with Superqubits}, {\tt arXiv:1001.3753}.










\bibitem{182} P. L\'evay,  G. S\'arosi, {\em Hitchin functionals are related to measures of
entanglement}, Phys. Rev. {\bf D86} (2012) 105038.


\bibitem{22}
S. James Gates, Jr., Kory Stiffler,  {\em  Adinkra `Color'
Confinement In Exemplary Off-Shell Constructions Of 4D, ${\cal N}$ =
2 Supersymmetry Representations}, { \tt arXiv:1405.0048}.

\bibitem{23}
 Y. X Zhang,  Adinkras for Mathematicians,  {\tt arXiv:1111.6055}




\bibitem{19}
M. A. Nielsen and I. L. Chuang,  {\em Quantum Computation and
Quantum Information},  Cambridge University Press, New York, NY,
USA, 2000.



\bibitem{20}

D. R. Terno, {\em  Introduction to relativistic quantum
information}, {\tt arXiv:quant-ph/0508049}.

\bibitem{21}

M. Kargarian, {\em Entanglement properties of topological color
codes}, Phys. Rev. {\em A 78}  (2008)062312, {\tt arXiv:0809.4276}.




\bibitem{24}
B. L. Douglas, S. James Gates Jr., Jingbo B. Wang,  {\em
Automorphism Properties of Adinkras}, arXiv:1009.1449


\bibitem{25}
C.F. Doran, M.G. Faux, S.J. Gates, Jr., T. Hubsch, K.M. Iga, G.D.
Landweber, {\em  Adinkras and the Dynamics of Superspace
Prepotentials}, {\tt  arXiv:hep-th/0605269}.

\bibitem{26}

 C.F. Doran, M.G. Faux, S.J. Gates, Jr., T.
Hubsch, K.M. Iga, G.D. Landweber, {\em  On Graph-Theoretic
Identifications of Adinkras, Supersymmetry Representations and
Superfields}, Int.J.Mod.Phys.{\bf A22}(2007)869-930,2007, {\tt
arXiv:math-ph/0512016}.

\bibitem{27}

 M. Faux, S. J. Gates Jr, {\em Adinkras: A Graphical Technology for
Supersymmetric Representation Theory } Phys.Rev. {\bf D71} (2005)
065002, {\tt  arXiv:hep-th/0408004}.

\bibitem{28}

 C. Doran, K. Iga, G. Landweber, {\em An
application of Cubical Cohomology to Adinkras and Supersymmetry
Representations},  {\tt arXiv:1207.6806}.


\bibitem{29} A. Salam and E. Sezgin, \emph{Supergravities in diverse
Dimensions}, Edited by A. Salam and E. Sezgin, North--Holland, World
Scientific \textbf{1989}, vol.1.


\bibitem{290} K. Intriligator and N. Seiberg, Phys. Lett. {\bf B387} (1996) 513.


\bibitem{291} K. Hori, H. Ooguri and C. Vafa, Nucl. Phys. {\bf B504} (1997) 147.

\bibitem{30}
  E. Witten, {\em Notes On Supermanifolds and
Integration},  {\tt arXiv:1209.2199}.
\bibitem{31}
 M. Aganagic, C. Vafa, {\em Mirror
symmetry and supermanifolds},  {\tt hep-th/0403192}.


\bibitem{32}
 A. Belhaj, L.B. Drissi, J. Rasmussen, E.H.
Saidi, A. Sebbar, {\em  Toric Calabi-Yau supermanifolds and mirror
symmetry}, J.Phys. {\bf A38} (2005) 6405-6418, {\em
arXiv:hep-th/0410291}.

\bibitem{320}
T. Prudencio, D. J. Cirilo-Lombardo, E. O. Silva, H. Belich, Black
hole qubit correspondence from quantum circuits, e-Print:
arXiv:1401.4196 [quant-ph].

\bibitem{321}
B.R. Greene, String Theory on Calabi Yau Manifolds, hep-th/9702155.

\bibitem{322}

 P. Aspinwall, K3 surfaces and String Duality, hep-th/961117
\bibitem{323}

 M. Rios, { \em
Extremal Black Holes as Qudits}, {\tt arXiv:1102.1193}.








\end{thebibliography}
\end{document}